\documentclass[11pt,a4paper]{article}

\usepackage[utf8]{inputenc}
\usepackage{amsmath}
\usepackage{algorithm}
\usepackage{algpseudocode}
\usepackage{amsfonts}
\usepackage{amssymb}
\usepackage{physics}
\usepackage{float}
\usepackage{geometry}
\usepackage{cleveref}
\usepackage{graphicx,color}
\usepackage[numbers,sort&compress]{natbib}
\usepackage[noblocks]{authblk}

\AtBeginEnvironment{align}{\setcounter{subeqn}{0}}
\newcounter{subeqn}

\renewcommand{\d}{\text{d}}

\begin{document}

\title{Learning the Kohn--Sham map with neural operators for quasi-linear scaling density functional theory}

\author[1$\dagger$]{Danish Khan}
\author[1$\dagger$]{Maurice D. Hanisch}
\author[1,3]{Nikolai Argatoff}
\author[1]{Evan Xie}
\author[2,4]{Sandeep Sharma}
\author[1*]{Anima Anandkumar}

\makeatletter
\@namedef{@sep3}{\Authsep\authorcr}
\makeatother

\affil[1]{Department of Computing and Mathematical Sciences,
California Institute of Technology}

\affil[2]{Division of Chemistry and Chemical Engineering,
California Institute of Technology}

\affil[3]{Department of Mathematics,
ETH Z\"urich}

\affil[4]{Marcus Center for Theoretical Chemistry, Pasadena, CA 91125, USA}
\affil[*]{Corresponding author. Email: anima@caltech.edu}

\affil[$\dagger$]{\bf Equal contribution.}

\maketitle

\begin{abstract}
Kohn--Sham density functional theory (DFT) underpins electronic-structure simulations, but repeated orbital diagonalizations lead to cubic scaling, restricting quantum calculations to modest scales only.
Eliminating these auxiliary orbitals while retaining Kohn--Sham accuracy is the central goal of orbital-free DFT, but both analytical and machine-learning methods have so far fallen short.
Prior learning approaches either try to learn the variational kinetic-energy functionals, which are ill-conditioned, or directly predict the ground state, which extrapolate poorly to larger systems.
Instead, we identify the Kohn--Sham map as the right learning target for orbital-free DFT. 
It maps a Kohn--Sham potential directly to the corresponding density and noninteracting kinetic energy, quantities otherwise obtained through an orbital diagonalization.
Focusing on the density component in this work, a domain-invariant $\mathrm{SE}(3)$-equivariant 
Fourier neural operator learns to predict it from the potential as input on real-space grids, enabling stable quasi-linear scaling SCFs.
Trained jointly on 8,504 molecules and solids, a single model generalizes to out-of-distribution organic molecules, insulators, and metals. 
For the first time, the same method converges SCFs across these systems without explicitly constructing Kohn--Sham orbitals, while reproducing densities, electronic spectra, and structural observables at Kohn--Sham DFT accuracy.
Linear-scaling SCFs additionally allow converging magnesium dislocation densities containing up to 82,500 valence electrons on a single GPU.

\end{abstract}

\section{Introduction}

More than six decades after Hohenberg and Kohn established the ground-state density as the fundamental variable of electronic structure, extending predictive first-principles calculations to mesoscopic scales remains a central challenge~\cite{hohenberg1964inhomogeneous}. Kohn and Sham made density functional theory (DFT) computationally practical, leading to its ubiquitous use across chemistry, materials science, condensed-matter physics, and biophysics~\cite{kohn1965self,becke2014perspective,jones2015density,huang2023central}. Their formulation introduces an auxiliary system of non-interacting electrons whose orbitals must be solved for and orthogonalized at every self-consistent field (SCF) iteration. 
These operations scale nominally as $\mathcal O(N^3)$ with $N$ electrons, restricting the length scales accessible to first-principles simulations of extended defects, interfaces, electrochemical environments, and biological systems~\cite{Cole_2016,das_fast_2019}.

The diagonalization of Kohn--Sham Hamiltonians is among the most frequently 
repeated and computationally consequential eigenvalue problems in modern science.
As an example, nearly 30\% of the workload at the US National
Energy Research Scientific Computing Center in 2018 was attributed to DFT
calculations~\cite{pederson2022large}.

The search for linear-scaling electronic structure methods has a long history.
Several efforts leverage Kohn's nearsightedness principle to truncate the density matrix (DM) through localized orbitals, leading to sparse-matrix operations, while others use DM purification methods to impose idempotency
~\cite{kohn1996density,bowler2012methods}.
Other approaches include Fermi-operator expansions, selected inversion, and stochastic
trace estimation~\cite{lin2013accelerating,baer2013self}.
These methods extend
the reach of KS-DFT, but their performance depends on density-matrix decay,
sparsity, dimensionality, or stochastic error. 
There is therefore no general framework for near-linear scaling DFT with KS accuracy across molecules, metals, insulators,
and chemically heterogeneous systems.

Hohenberg and Kohn originally formulated DFT as a variational theory of the
density alone. 
At fixed spatial resolution, the size of the density
representation grows linearly with system size, offering a route to linear or
quasi-linear scaling assuming a similar cost of applying the energy functionals. Orbital-free DFT
(OF-DFT) seeks to recover this density-only formulation by minimizing the
energy directly over the density~\cite{mi2023orbital}. 
Its central unknown is
the non-interacting kinetic energy density functional (KEDF). 
KS-DFT evaluates this quantity from the
auxiliary orbitals, whereas OF-DFT requires an explicit density functional
$T_s[n]$. Existing analytical KEDFs are efficient but lack the accuracy and
transferability needed to describe shell structure, chemical bonding, and
nonlocal response across molecules and materials
~\cite{mi2023orbital}. 
Machine-learned KEDFs have improved this description
~\cite{zhang2024overcoming,chen2026machine}, but stable self-consistent use
requires accurate functional derivatives, which are considerably harder to
learn than energies evaluated on prescribed densities
~\cite{remme2025stable}.
OF-DFT thus defines the goal, namely a
density-only solver, but leaves open the best object to learn.

In contrast to OF-DFT, models that infer a converged density from atomic structure~\cite{brockherde2017bypassing,jorgensen2022equivariant,li2025image} take the direct approach of learning the composite ground-state (GS) map $\mathcal G_{\mathrm{GS}}:v_{\mathrm{ext}}\mapsto n^\star$. 
Here, $v_{\mathrm{ext}}$ and $n^\star$ denote the ionic external Coulomb potential and corresponding ground-state density, respectively. 
In exact theory, this map involves the interacting many-electron ground-state problem. KS-DFT determines $n^\star$ through a sequence of simpler noninteracting problems: each SCF iteration solves for the density associated with a KS potential and uses that density to construct the next potential, allowing later iterations to refine earlier estimates. 
A direct GS prediction model must instead learn the endpoint of an arbitrarily long, XC-specific trajectory. 
A similar direct endpoint-learning strategy underlies machine-learned interatomic potentials, which learn the energy map $v_{\mathrm{ext}}\mapsto E[n^\star]$~\cite{unke2021machine}. Alternatives that use near-converged quantum-mechanical (from e.g. semi-empirical calculations) features as input rather than being learned directly like Orbitall~\cite{kang2025orbitall} greatly simplify this map and have been shown to outperform state-of-the-art MLIPs.
The benefit of retaining intermediate computation is also evident through inference-time reasoning in large language models, where difficult answers are constructed through intermediate steps rather than in one prediction~\cite{wei2022chain}.

Models that predict the converged DFT Hamiltonian from atomic structure are
also direct ground-state models
~\cite{li2022deep,gong2023general,kaniselvan2025learning}. 
In KS-DFT, the Hamiltonian can be directly constructed from the density alone.
Predicting it separately therefore introduces an unnecessary basis-dependent matrix whose number of entries grows quadratically with the system size.
NeuralSCF~\cite{song2026neural} learns the composite update from the density at a KS-DFT SCF iteration and atomic structure to the next density, and iterates this map to self-consistency, outperforming a matched one-shot ground-state predictor. 
Because the KS potential is not supplied explicitly, however, the model must learn both its XC-dependent construction, non-local Hartree screening, and then the subsequent non-interacting solution. 
Additionally, it depends on the XC approximation used to generate the learned SCF trajectory, as well as the basis representation used for the density, limiting its transferability between molecules and periodic systems.

\begin{figure}[!htbp]
    \centering
    \includegraphics[width=0.95\textwidth]{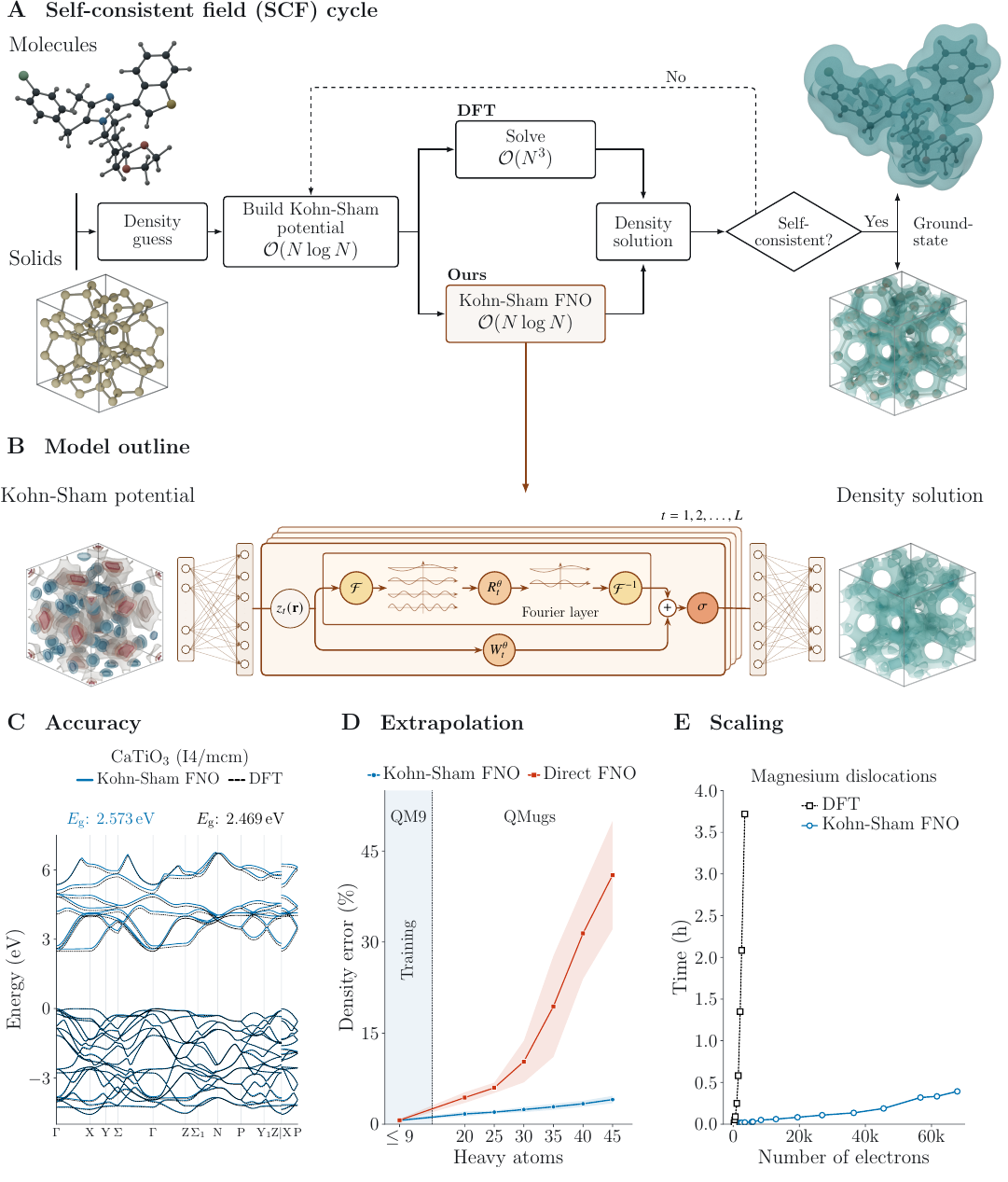}
    \caption{\textbf{Summary of the Kohn--Sham Fourier Neural Operator (Kohn-Sham FNO) based self-consistent field (SCF) cycles for density optimization.}
    (A) Conventional DFT and Kohn-Sham FNO workflows.
    The model replaces the orbital-solution step (see SI) while retaining the rest of a standard DFT SCF cycle.
    Our SCF implementations for both molecules and solids closely follow Quantum ESPRESSO (QE) 7.5~\cite{giannozzi2009quantum, giannozzi2017advanced} with minor modifications for retaining quasi-linear scaling.
    All DFT calculations are also performed using QE.
    See SI.
    (B) Schematic of the model mapping an input potential to the density produced by solving the corresponding Kohn-Sham equations.
    See SI.
    (C) Band structures for tetragonal perovskite CaTiO$_3$ (MC3D~\cite{huber2026mc3d} ID \texttt{mc3d-69107}; $I4/mcm$, No.~140)
    obtained using the Kohn-Sham FNO (blue) converged densities and regular DFT (black).
    $E_g$ values denote the corresponding band gaps. 
    See SI.
    (D) Density error versus molecular size for self-consistent and direct
    ground-state prediction on small-organic QM9~\cite{ramakrishnan2014quantum} and larger drug-like QMugs molecules~\cite{isert2022qmugs}.
    See Fig.~\ref{fig:fig3_mols}.
    (E) Measured total SCF cycle wall time versus electron count for QE and Kohn-Sham FNO. See SI.
    }
    \label{fig:workflow}
\end{figure}

\newpage
{\bf Our Approach:} A natural and universal learning target already exists in every Kohn--Sham calculation. 
At each SCF iteration, the density $n$ defines an effective potential $v_{\mathrm{KS}}[n]$. 
Solving its non-interacting Hamiltonian and occupying the eigenstates produces the output density,
\begin{equation}
\left[
-\frac{1}{2}\nabla^2
+
v_{\mathrm{KS}}[n](\mathbf r)
\right]
\phi_p(\mathbf r)
=
\varepsilon_p\phi_p(\mathbf r)
\quad\Longrightarrow\quad
n^{\mathrm{out}}(\mathbf r)
=
\sum_p f_p
\left|\phi_p(\mathbf r)\right|^2 
\label{eq:ks_intro}
\end{equation}
The SCF converges when $n^{\mathrm{out}}=n$, but every preceding iteration
also supplies an exact sample of the same forward universal operator,
\begin{equation}
\mathcal{G}_{\mathrm{KS}}^{(n,T)}:
v_{\mathrm{KS}}[n](\mathbf r)
\longmapsto
\left(
n^{\mathrm{out}}(\mathbf r),
T_s[n^{\mathrm{out}}]
\right),
\label{eq:ks_operator_intro}
\end{equation}
where the electron number, boundary conditions, and occupations are implicit.
Both the density and kinetic energy are therefore one non-interacting solve away from a prescribed KS potential, and together form a complete orbital-free DFT target.

Stable density optimization has historically been the harder test for OF-DFT
~\cite{mi2023orbital,remme2025stable} and hence in this work we learn only the density component, i.e. $v_{\mathrm{KS}}[n](\mathbf r)
\longmapsto
n^{\mathrm{out}}(\mathbf r),$
denoted $\mathcal G_{\mathrm{KS}}$ for performing stable orbital-free SCFs at inference.
Consequently, we obtain total energies and electronic spectra via a fixed-density post-SCF diagonalization in the following results.

This forward formulation is the inverse of the relation used in traditional OF-DFT.
For a local KS potential, minimization of the total energy at fixed electron number gives
\begin{equation}
\left.
\frac{\delta T_s[\rho]}{\delta \rho(\mathbf r)}
\right|_{\rho=n^{\mathrm{out}}}
=
\mu-v_{\mathrm{KS}}[n](\mathbf r),
\label{eq:kedf_inverse_intro}
\end{equation}
where $\mu$ enforces the electron number. 
A KEDF therefore maps the density to
the kinetic energy, and its functional derivative recovers the generating effective potential up to
the additive constant $\mu$ : $n^{\rm out}(\mathbf{r}) \mapsto \mu - v_{\rm KS}[n](\mathbf{r})$, which is the inverse of the eq.~\ref{eq:ks_operator_intro} operator. By contrast, the two
outputs in Eq.~\ref{eq:ks_operator_intro} map the generating potential directly
to its density and kinetic energy.

Compared to direct prediction models, $v_{\mathrm{KS}}[n]\mapsto n^{\rm out}$ is a more elementary map defined via a single non-interacting Hamiltonian solution.
Compared to MLIPs, a simpler forward
counterpart is $v_{\mathrm{KS}}[n^\star]\mapsto T_s[n^\star]$, after which the
remaining energy terms are explicit. 
Section~\ref{sec:results_ofdft} discusses
this energy analogy further, while Fig.~\ref{fig:fig3_mols} directly compares forward KS and
one-shot density prediction.

Equation~\ref{eq:ks_operator_intro} gives the local form of the KS operator while
our calculations use
nonlocal pseudopotentials~\cite{payne1992iterative,hamann2013optimized} to smooth the core region and allow us to use uniform-grid
fast Fourier transforms (FFT). 
For each
fixed set of nonlocal projectors, we combine their operator with the kinetic
energy in a generalized non-interacting kinetic functional defined by
constrained search~\cite{levy1979universal,lieb1983}. 
This preserves the joint
forward map and its inverse Euler relation for the pseudopotential problem.
See SI.
This defines one operator across SCF iterations, 
chemical composition, geometry, and exchange-correlation (XC)
approximations. 
These inputs modify the KS potential but not the non-interacting solution operator being learned. 
Potential-density pairs from
different XC approximations can therefore be combined in training. 
We test this separation
by applying a PBE-trained model~\cite{pbe} to PBEsol
SCFs~\cite{perdew2008restoring} without retraining
[Fig.~\ref{fig:fig4_crystal_props}].

The KS map is an operator between spatial fields, making neural operators a
natural model class for this framework. Neural operators learn maps between function
spaces rather than between fixed-dimensional vectors.
We specifically utilize Fourier neural operator (FNO) since the Fourier layers
of an FNO capture global nonlocality with quasi-linear
$\mathcal O(N_g\log N_g)$ scaling
~\cite{kovachki2023neural,azizzadenesheli2024neural} on the same grid used to construct the Kohn-Sham potential.
The complete update therefore scales as $\mathcal O(N_g\log N_g)$ for $N_g$
grid points.
Alternative architectures, e.g., graph neural networks rely either on finite spatial cutoffs or on globally coupled operations whose cost grows quadratically with system size, e.g., transformers.

In this work, we introduce a new domain-invariant, $\mathrm{SE}(3)$-equivariant FNO architecture.
Unlike the standard FNO, whose spectral weights are tied to a fixed domain
size, the domain-invariant FNO evaluates radially factorized filters as functions of
physical reciprocal-space radius. The same learned filters can therefore be
sampled on the reciprocal-space grids of differently sized domains, allowing
one architecture to treat molecules and materials across system sizes. The
radial filters and spherical mode truncation make the spectral convolution
rotation equivariant; together with its translation equivariance, this yields
full $\mathrm{SE}(3)$ equivariance.

We first show that learning the KS-map leads to significantly improved stability, and chemical extrapolation than the inverse OF-DFT and direct ground-state learning approaches respectively.
Following this, we show, for the first time, a single method converging SCF calculations without invoking the auxiliary Kohn-Sham orbitals across organic molecules of varying sizes as well as insulating and metallic systems spanning the first 5 rows of the periodic table.
The corresponding converged densities are shown to produce observables nearly indistinguishable in quality from a regular KS-DFT calculation.
This is verified using electronic and structural observables, including band structures, band gaps, density of states, equation of state curves, equilibrium volumes, and bulk moduli. 
The computational advantage over conventional KS-DFT is particularly large
for periodic systems since the
Kohn--Sham FNO inference loop evolves only the $\mathbf{k}$-independent,
lattice-periodic potential and density on the unit cell and requires no explicit
$\mathbf{k}$-point sampling~\cite{shao2021dftpy} skipping $N_{\rm iterations} \times N_{\mathbf{k}}$ diagonalizations. 
Finally, after training on cells with at most 364 atoms, we obtain stable SCFs for
magnesium dislocations containing up to 8,250 atoms on one GPU, empirically verifying
the expected near-linear scaling.

\section{Results \& Discussion}

Unless otherwise stated, results are obtained using a single domain-invariant,
radially factorized FNO trained on density-potential pairs from
2,004 molecules and 6,500 solids at their equilibrium geometries
(Fig.~\ref{fig:fig3_mols}A). 
When the domain-invariant FNO learns the KS solution operator and
is embedded in the SCF cycle, we refer to the resulting model $\&$ method interchangeably as Kohn-Sham FNO for brevity. 
At inference, the model replaces the KS eigensolver within a fixed-point SCF iteration to optimize the densities. 
Fig.~\ref{fig:workflow} summarizes the SCF workflow and model architecture, with details of the training data, architecture, optimization procedure, and quasi-linear scaling SCF implementation. See SI.

\begin{algorithm}[!htbp]
\caption{\textbf{Self-consistent field cycle with Kohn--Sham FNO.}}
\label{alg:ksfno_scf}
\small
\noindent\textbf{Input:} Atomic structure, convergence tolerance $\varepsilon$, and maximum iteration count $I_{\max}$.
\vspace{0.35em}

\noindent
\begin{tabular}{@{}r@{\hspace{0.65em}}p{\dimexpr\linewidth-2.35em\relax}@{}}
1 & Construct $v_{\rm ext}^{\rm loc}$ and $n_{\rm SAD}$. Initialize $n_0=n_{\rm SAD}$ and the mixing factor $\alpha_0$. \\[1pt]
2 & \textbf{for} $i=0,\ldots,I_{\max}-1$ \textbf{do} \\[1pt]
3 & \quad Construct $v_{\rm KS}=v_{\rm ext}^{\rm loc}+v_{\rm H}[n_i]+v_{\rm xc}[n_i]$. \\[1pt]
4 & \quad Predict $\widetilde n_i^{\rm FNO}=n_{\rm SAD}+F_\theta[v_{\rm KS},v_{\rm ext}^{\rm loc}]$ and project $n_i^{\rm FNO}=\mathcal P_{N_e}[\widetilde n_i^{\rm FNO}]$. \\[1pt]
5 & \quad Evaluate the relative fixed-point residual $R_i^{\rm rel}=\frac{1}{N_e}\!\int |n_i^{\rm FNO}(\mathbf r)-n_i(\mathbf r)|\,\d\mathbf r$. \\[1pt]
6 & \quad \textbf{if} $R_i^{\rm rel}<\varepsilon$ \textbf{then return} $n^\star=n_i^{\rm FNO}$. \\[1pt]
7 & \quad Mix to obtain a new density $n_{i+1}=\mathcal P_{N_e}[(1-\alpha_i)n_i+\alpha_i n_i^{\rm FNO}]$. \\[1pt]
8 & \textbf{end for} \\[1pt]
9 & \textbf{return} unconverged
\end{tabular}

\vspace{0.35em}
\noindent\textbf{Output:} A self-consistent density $n^\star$, potential $v_{\rm KS}[n^\star]$, and Hamiltonian $h_{\rm KS}[n^\star]$ or an explicit convergence failure.
\end{algorithm}

\subsection{Stable optimization \& improved extrapolation through the forward map}
\label{sec:results_ofdft}

\begin{figure}[!htbp]
    \centering
    \includegraphics[width=\textwidth]{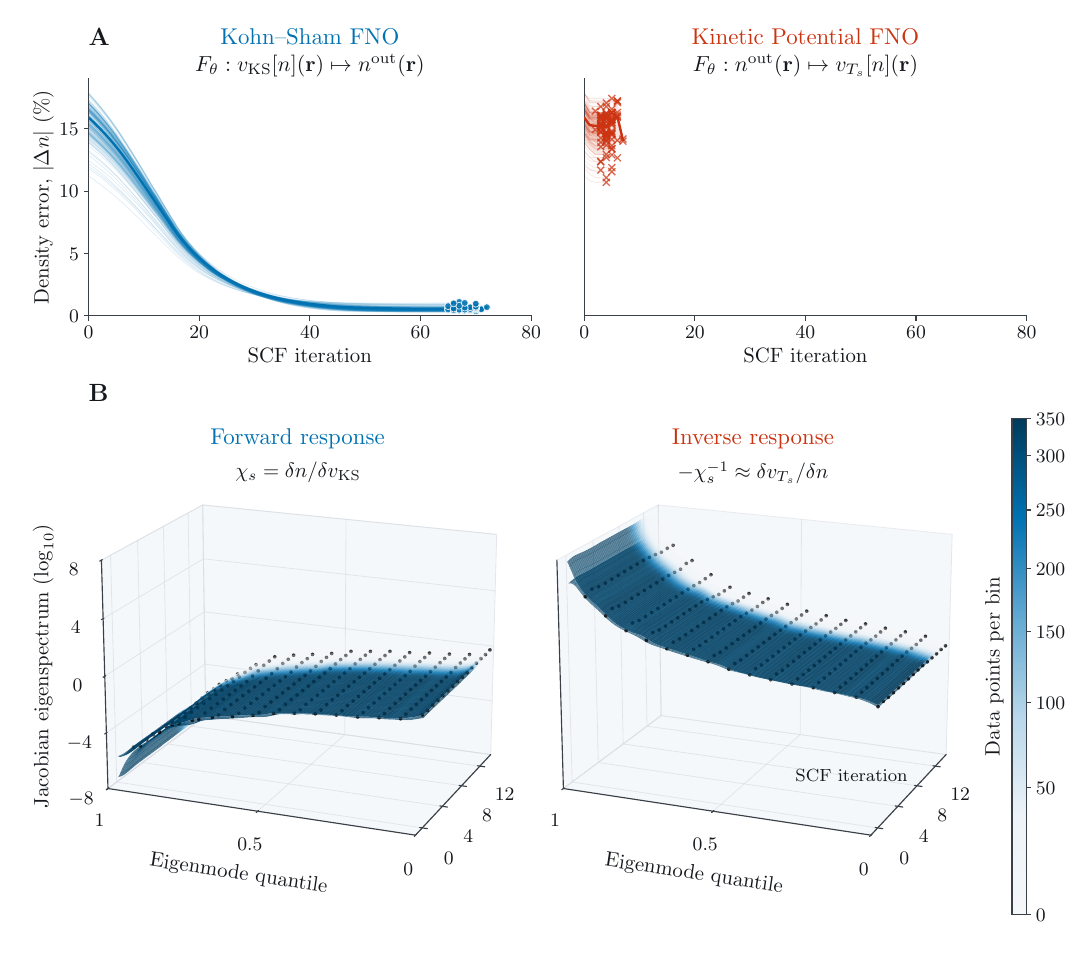}
    \caption{\textbf{Learning the forward orbital-free KS map avoids inverse-response
    instability.}
    (A) Density error relative to PBE~\cite{pbe} ground-states for 100 held-out small-organic molecules from the QM9 dataset~\cite{ramakrishnan2014quantum}. See SI.
    Kohn--Sham FNO (blue) maps the KS potential to its output
    density and is iterated to a fixed-point. 
    The kinetic-potential FNO (red) uses the same domain-invariant FNO model as the backbone and learns the density to kinetic-potential relation and is used to solve the Euler equation based on the implementation in DFTpy~\cite{shao2021dftpy}. 
    See SI.
    Both models use the same training data (Fig.~\ref{fig:fig3_mols}A) and FNO
    backbone. 
    Thin curves show individual molecules, bold curves show their
    mean, and circles and crosses mark converged and failed calculations,
    respectively.
    (B) Binned eigenspectra of the projected forward response $\chi_s$ (left)
    and inverse response $-\chi_s^{-1}$ (right) over 1,100 PBE
    SCF states for the same molecules.
    Modes are ordered from strongest to
    weakest forward response; points mark bin medians and surface color gives
    the number of eigenvalues per bin.
    The responses were computed using all-electron calculations in Gaussian basis sets with the PySCF~\cite{sun2018pyscf} and KS-PIES~\cite{nam2021ks} packages.
    See SI.
    }
    \label{fig:fig2_ofdftcomp}
\end{figure}

The learning target determines both the numerical problem presented to the
model and how its errors enter the final density.
We compare three targets that seek to eliminate the orbital bottleneck: the forward KS map used herein,
its inverse kinetic-potential map used in traditional orbital-free density optimizations,
and direct prediction of the ground-state density from the atomic environment.
The forward KS map lies between the other two. 
It neither inverts the potential-to-density response nor compresses the complete nonlinear ground-state problem into a single prediction. 
We first test the consequences of this reformulation for density-optimization stability and then for chemical
generalization.

For a non-interacting $v$-representable density, the Euler equation identifies the kinetic potential as $v_{T_s}[n^{\mathrm{out}}]=\mu-v_{\mathrm{KS}}[n]$. See SI.
Learning a kinetic potential therefore reverses the same relation learned by the Kohn--Sham FNO. 
The distinction is consequential.
The forward map is defined by a one-particle ground-state problem for every KS potential. 
Its inverse is defined only for non-interacting $v$-representable densities. 
Recovering the corresponding potential generally requires an iterative inversion such as the Wu-Yang
method~\cite{wu2003direct}, and a complete set of constraints characterizing the
$v$-representable domain is unknown~\cite{kohn1983v,trushin2024violations}.
Trial densities generated during orbital-free optimization can therefore 
leave the $v$-representable domain sampled by exact inverse labels.
A scalar kinetic-energy functional is subject to the same issue when used
variationally. 
Its forward counterpart is the explicit map
\begin{equation}
v_{\mathrm{KS}}[n](\mathbf r)
\longmapsto
T_s[n^{\mathrm{out}}],
\label{eq:results_forward_energy_map}
\end{equation}
where $n^{\mathrm{out}}$ is the density produced by the input
potential. 
The density map
$v_{\mathrm{KS}}[n]\mapsto n^{\mathrm{out}}$ and the energy map in
Eq.~\ref{eq:results_forward_energy_map} therefore follow from the same single
one-particle diagonalization. 
The pseudopotential calculations below obey the
same relations with $T_s$ replaced by the generalized functional
$\widetilde T_s$. See SI.

We isolate the effect of mapping direction by training a kinetic-potential
FNO with the same backbone architecture and the same training data as the
Kohn--Sham FNO. 
The forward model learns $v_{\mathrm{KS}}[n]\mapsto n^{\mathrm{out}}$, whereas the inverse model
learns $n^{\mathrm{out}}\mapsto\mu-v_{\mathrm{KS}}[n]$, using the
generalized kinetic potential of the nonlocal-pseudopotential formulation. See SI.
Starting from the same
superposition-of-atomic-densities (SAD) guess, we apply the forward model in a
fixed-point SCF procedure and the inverse model in a potential-only optimization. See SI.
The resulting trajectories, measured against the PBE reference density using the density-error metric, are qualitatively different [Fig.~\ref{fig:fig2_ofdftcomp}A]. See SI.
All forward-model calculations reach a
self-consistent fixed point, progressively reducing the density error with
respect to the reference PBE density from approximately $10$--$18\%$ for the
initial guess to less than $1\%$. 
By contrast, the inverse-model optimizations fail after only a few iterations: the learned Euler residual
does not provide a descent direction for which the backtracking procedure can
continue, and the density remains close to its initial error. 
This comparison uses the same training data, model architecture, as well as equivalent training and inference strategies; its primary
difference is whether the KS relation is learned in the forward or inverse direction.

The same difficulty arises when a learned energy functional is differentiated
to optimize the density. 
Remme \textit{et al.} obtained stable QM9
SCFs by training the combined kinetic and XC energy
$E_{\mathrm{TXC}}[n]$ on about 107,000 molecules and 21 perturbed SCF states
per molecule, giving roughly 2.25 million training labels
~\cite{remme2025stable}. 
In their ablation, ordinary SCF training data led to
a 28\% convergence-failure rate, compared with no failures using the perturbed
data. 
Kohn--Sham FNO is only trained on 59,500 total ordinary SCF training labels from 8,504
molecular and periodic structures, while converging SCFs in both domains. 
Although the two
architectures are not directly comparable, the large contrast shows how much
off-equilibrium data may be needed to learn the inverse density-to-potential
map stably.

The stability difference between the two models stems from how the two learned maps are conditioned. 
A linear perturbation of the KS potential produces
$\delta n=\chi_s\delta v_{\mathrm{KS}}$, where
$\chi_s=\delta n/\delta v_{\mathrm{KS}}$ is the non-interacting density
response. See SI.
On the particle-number conserving subspace, linearization of the
Euler equation instead gives
$\delta v_{T_s}/\delta n=-\chi_s^{-1}$. 
We evaluated the spectrum of
$\chi_s$ independently for every recorded SCF iteration of the same 100 QM9 molecules
using all-electron KS response calculations.
See SI.
The forward spectrum in
Fig.~\ref{fig:fig2_ofdftcomp}B contains many weak-response modes with small
$|\lambda_j(\chi_s)|$. 
Such modes suppress potential errors in the forward
map, but inversion turns them into sensitivities proportional to
$|\lambda_j(\chi_s)|^{-1}$ and amplifies density errors by several orders of
magnitude. 
The mirrored spectra persist throughout the SCF trajectories, so
the inverse conditioning problem is not confined to the converged density.
It applies whether the kinetic potential is learned directly, as in our
controlled inverse model, or obtained by differentiating a learned
kinetic-energy functional.
\\
\begin{figure}[!htbp]
    \centering
    \includegraphics[width=\textwidth]{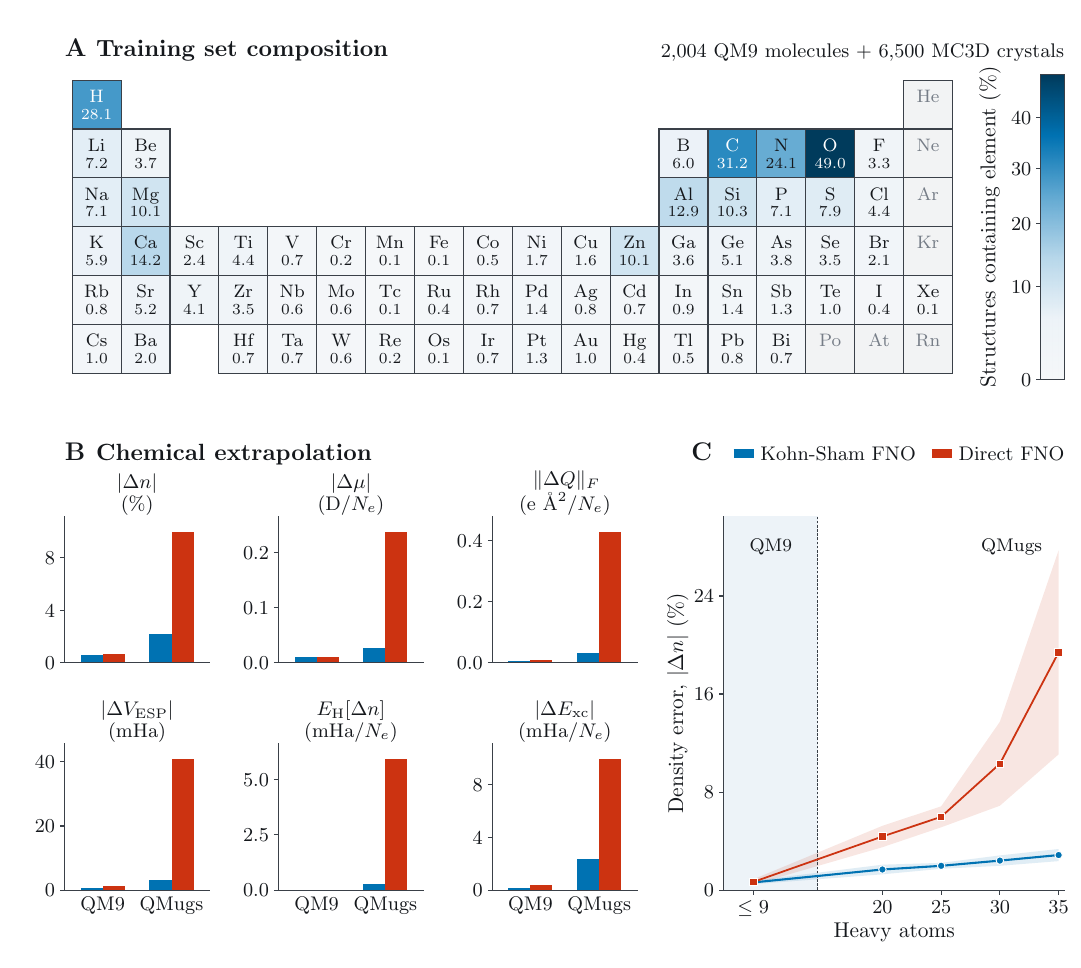}
    \caption{\textbf{Improved extrapolation via the self-consistent Kohn--Sham map.}
    (A) Elemental composition of the shared training set of 2,004 QM9
    molecules~\cite{ramakrishnan2014quantum} and 6,500 MC3D crystals~\cite{huber2026mc3d}.
    Each value is the percentage of
    structures containing that element; gray denotes absence from training.
    (B) Mean errors of the self-consistent Kohn--Sham FNO (blue) densities and direct ground-state predicted densities on 100 held-out QM9 molecules containing the elements C, H, N, O, F and 200 larger drug-like molecules from the QMugs dataset~\cite{isert2022qmugs} containing the elements C, H, N, O, F, S, Cl, P.
    From left to right, the metrics are the normalized $L^1$
    density error, Euclidean norm of the dipole-vector error, Frobenius norm of the quadrupole-tensor error, electrostatic-potential error,
    Coulomb-weighted density-error metric, and
    absolute XC-energy error; quantities labeled $/N_e$ are normalized per
    valence electron.
    (C) Density error versus number of heavy atoms. 
    The QM9 bin at $\leq$9 heavy atoms consists of the same 100 test molecules from Fig.~\ref{sec:results_ofdft}.
    The QMugs bins from $20-35$ heavy atoms each consist of 50 randomly sampled ground-state conformers.
    All errors are relative to converged Quantum ESPRESSO PBE calculations.
    Data construction, the direct model, and the density metric: see SI.
    }
    \label{fig:fig3_mols}
\end{figure}
We next compare the forward operator with the opposite extreme: predicting
the converged density directly from the ionic environment.
Direct prediction models can
be accurate and transferable when symmetry, locality, and suitable density
representations are built into the architecture
~\cite{brockherde2017bypassing,grisafi2019transferable,lewis2021learning,jorgensen2022equivariant,li2025image}.
We again hold the FNO backbone, grid representation, and training structures
fixed and change the target. 
The direct model learns $v_{\mathrm{ext}}^{\mathrm{loc}}\mapsto n_{\mathrm{Ref}}^\star$ in one
evaluation, whereas the Kohn--Sham FNO learns one non-interacting solution
and recovers $n_{\mathrm{Ref}}^\star$ through self-consistency.

This distinction separates what must be learned from what can be evaluated
explicitly. 
Direct prediction must encode the entire PBE fixed point, which involves constructing the ground-state $v_{\rm KS}[n^{\star}]$ first via an arbitrarily long non-linear SCF trajectory involving Hartree and XC feedback and long-range charge redistribution. 
The Kohn--Sham FNO instead receives the reconstructed
Hartree and XC potentials at every iteration and repeatedly applies the same
one-particle solution operator. 
An error in one update changes the next input
potential and can be corrected by subsequent updates, while density mixing
damps unstable steps. 
Moreover, each conventional SCF trajectory supplies
multiple exact potential-density training pairs for the forward operator at
no additional electronic-structure cost, rather than only its final density.

On held-out QM9 molecules drawn from the same small-organic chemical space as
the molecular training set, both approaches are accurate
[Fig.~\ref{fig:fig3_mols}B, left]. 
The Kohn--Sham FNO and direct FNO obtain
density errors of $0.625\%$ and $0.662\%$, respectively, with dipole errors of
$0.010$ and $0.011$~D per electron. 
Their quadrupole,
electrostatic-potential, Hartree-energy, and XC-energy errors are similarly
close. 
The direct model therefore has sufficient capacity to represent the
ground-state map within its training distribution.
The difference emerges for QMugs drug-like molecules
~\cite{isert2022qmugs}, which are larger than QM9 and introduce S, Cl, and P
into molecular environments absent from the molecular training set
[Fig.~\ref{fig:fig3_mols}B, right].
Although these elements occur in the
solid-state portion of the shared training data, their bonding environments
and molecular domain sizes are out of distribution. 
Under this combined size, composition, and chemical-environment shift, the direct-model density
error rises to $9.97\%$, compared with $2.23\%$ for the self-consistent
Kohn--Sham FNO. 
The self-consistent cycle also reduces the dipole error from $0.237$ to
$0.026$~D per electron, the quadrupole error from $0.428$ to
$0.031$~e\,\AA$^2$ per electron, and the electrostatic-potential error from
$40.8$ to $3.28$~mHa.
The Hartree- and XC-energy errors fall from $5.91$ to
$0.267$~mHa per electron and from $9.92$ to $2.32$~mHa per electron,
respectively.
Figure~\ref{fig:fig3_mols}C shows that this extrapolation gap widens
systematically with molecular size: from 20 to 45 heavy atoms, the direct-model
density error increases from about $4\%$ to $41\%$, whereas the self-consistent
Kohn--Sham FNO grows only from about $1.5\%$ to $4\%$. This contrasting size
dependence is consistent with SCF feedback correcting intermediate errors
rather than requiring a one-shot model to extrapolate the complete density of
an increasingly large system.

The same idea applies to energy prediction. 
MLIPs learn the complete
ground-state map from the external potential represented by the atomic graph to the
ground-state energy~\cite{unke2021machine},
\begin{equation}
v_{\mathrm{ext}}
\longmapsto
E[n^\star].
\label{eq:direct_energy_map_results}
\end{equation}
Once an SCF has constructed the ground-state KS potential, the forward
alternative is
\begin{equation}
v_{\mathrm{KS}}[n^\star]
\longmapsto
\left(n^\star,T_s[n^\star]\right)
\longmapsto
E[n^\star],
\label{eq:forward_energy_map_results}
\end{equation}
where the external-potential, Hartree, XC, and nuclear contributions are evaluated
explicitly. 
The first map asks the model to infer the entire electronic
ground-state solution directly from the atoms; the second asks it to reproduce
one non-interacting KS solution after the self-consistent potential is already
known. 
The extrapolation advantage in Fig.~\ref{fig:fig3_mols} therefore
motivates the same hypothesis for energy prediction. 
We do not train a forward
energy model here, however, and leave this hypothesis for a 
future study involving a variationally consistent energy, density Kohn--Sham model.

Together, Figs.~\ref{fig:fig2_ofdftcomp} and~\ref{fig:fig3_mols} identify the
forward KS map as the favorable intermediate learning target among the three
formulations tested. 
Relative to the inverse kinetic-potential map, it avoids
inverse-response amplification and is queried only on potentials for which
the exact forward target is defined. 
Relative to direct ground-state prediction, it
factorizes a composite, XC-dependent fixed point into repeated applications
of an XC-independent one-particle solve with explicit physical feedback. 
In this sense, the KS map is the optimal target considered here: it removes the repeated eigensolution that dominates the cost while preserving the parts of KS-DFT that stabilize and transfer the calculation. 
This is not a claim of formal optimality over every possible representation, but it is the
only target in these controlled comparisons that combines stable 
self-consistent optimization with robust extrapolation.

\subsection{Periodic self-consistency without $\mathbf{k}$-points}
\label{sec:results_materials}

\begin{figure}[!htbp]
    \centering
    \includegraphics[width= \textwidth]{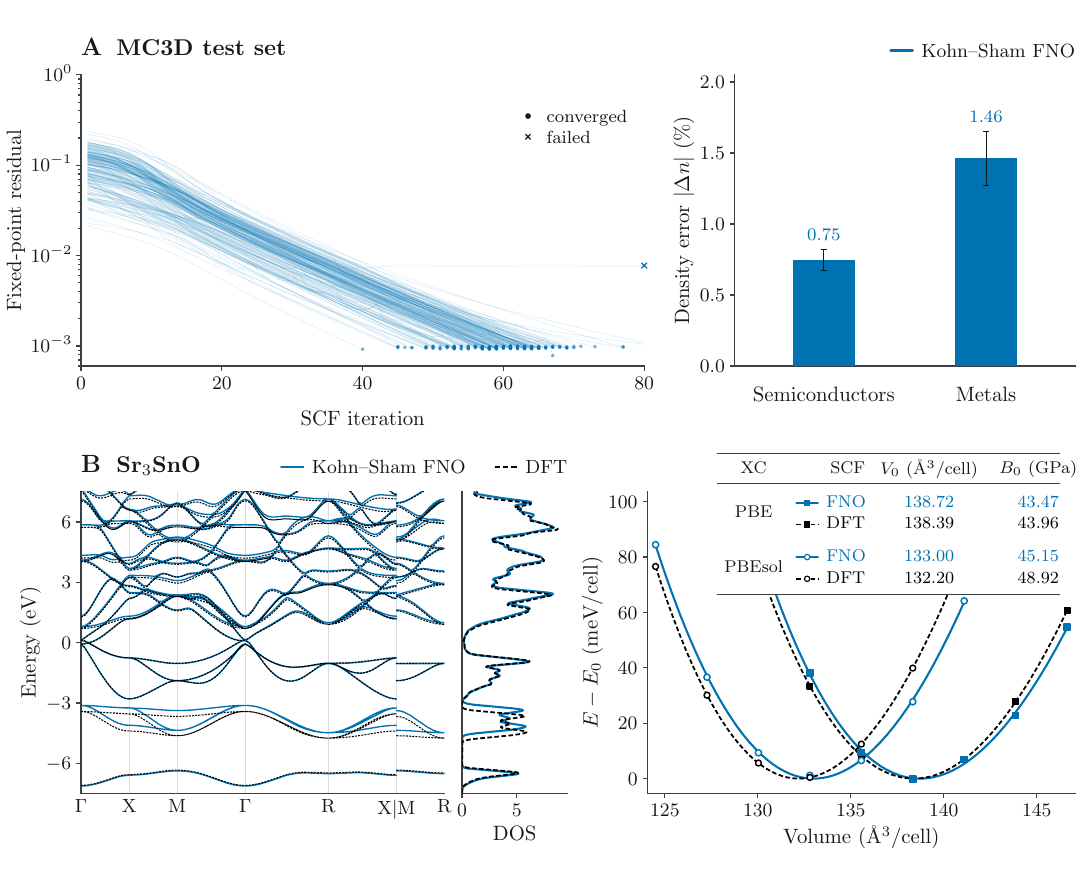}
    \caption{\textbf{Generalization across periodic systems, SCF trajectories, geometries and XC approximation.}
    (A) Fixed-point trajectories and final density errors for a held-out MC3D test set composed of 100 semiconductors and 100 metals as characterized by converged PBE calculations.
    Chemical composition of these crystals spans the same subspace of the periodic table as in Fig.~\ref{fig:fig3_mols}A.
    Thin curves denote individual crystals,
    dots and crosses mark converged and failed calculations, and bars summarize
    the mean errors for semiconductors and metals.
    (B) Results for metallic cubic Sr$_3$SnO (MC3D ID
    \texttt{mc3d-54098}; $Pm\overline{3}m$, No.~221). 
    Left, band structure
    and density of states from one fixed-density post-SCF diagonalization using
    the FNO density (blue) and from self-consistent PBE (black dashed), each
    referenced to its Fermi energy. 
    Right, clamped-ion PBE (squares) and
PBEsol~\cite{perdew2008restoring} (circles) equations of state; solid blue and dashed black curves
    denote FNO and DFT, respectively, and the table reports fitted equilibrium
    volumes $V_0$ and bulk moduli $B_0$. 
    The PBEsol test uses the PBE-trained
    model without fine-tuning and replaces only the XC potential in the KS potential construction during
    inference. See SI.
    Training, post-SCF observables, and EOS fitting: see SI.
    }
    \label{fig:fig4_crystal_props}
\end{figure}

Machine-learned orbital-free DFT has previously been applied to periodic
systems, but demonstrations have been limited to selected material classes
and have relied on local pseudopotentials constructed specifically for
orbital-free calculations~\cite{imoto2021order,sun2024machine}.
Here, we use the same Kohn--Sham FNO that drives the molecular SCFs in the
preceding section, trained jointly on molecular data and chemically diverse
metallic and semiconducting systems from MC3D~\cite{huber2026mc3d}.
See SI.
Its parameters, architecture, and inference procedure are unchanged across
molecules, metals, and semiconductors.
Moreover, it reproduces densities from
the transferable nonlocal norm-conserving pseudopotentials routinely used in
KS-DFT, whose orbital-dependent projectors cannot be treated directly within
conventional density-only orbital-free DFT.

The computational advantage over conventional KS-DFT is particularly large
for periodic systems. 
As in conventional periodic orbital-free DFT, the
Kohn--Sham FNO inference loop evolves only the $\mathbf{k}$-independent,
lattice-periodic potential and density on the unit cell and requires no explicit
$\mathbf{k}$-point sampling~\cite{shao2021dftpy}. 
The distinction lies in how the density update is
obtained. 
Conventional orbital-free DFT derives it from an approximate kinetic functional, whereas the Kohn--Sham FNO learns the composite KS operation that solves the Hamiltonian at every sampled $\mathbf {k}$-point and contracts the occupied states into the Brillouin-zone-summed density. 
A single model evaluation therefore replaces all $\mathbf{k}$-resolved diagonalizations in each conventional KS-DFT iteration while retaining the KS potential-density
feedback. See SI.

On 200 held-out MC3D crystals, the main model drives the fixed-point residual
smoothly toward zero, with all but one calculation converging within 80
iterations [Fig.~\ref{fig:fig4_crystal_props}A].  
The resulting mean density errors are
$0.75\%$ for semiconductors and $1.46\%$ for metals.

To test whether the converged FNO density also preserves observables that are
sensitive to the ground-state Hamiltonian, we consider the held-out metallic
Sr$_3$SnO cubic antiperovskite
(MC3D ID \texttt{mc3d-54098}, space group $Pm\overline{3}m$)
~\cite{huber2026mc3d}.
Its multiband electronic structure, with several dispersive states near the
Fermi level, provides a nontrivial test of both spectral and energetic
fidelity. 
One fixed-density post-SCF diagonalization using the FNO
density reproduces the self-consistent PBE band structure and density of
states [Fig.~\ref{fig:fig4_crystal_props}B]. 
At the reference volume, its
Harris total energy differs from self-consistent PBE by only
$11.4$~meV/cell ($2.29$~meV/atom).

The equation of state tests transfer beyond the equilibrium geometries used
for training. See SI.
Across seven clamped-ion
volumes spanning $V/V_{\rm ref}=0.94$--$1.06$, the PBE FNO and DFT energy
curves are nearly coincident: their fitted equilibrium volumes differ by
$0.23\%$ and their bulk moduli by $1.1\%$. 
Changing the volume presents a new
KS potential, but the learned task remains the same one-particle solution map
rather than direct prediction of a geometry-specific ground-state property.

We then change the XC approximation itself. 
PBEsol is a generalized gradient approximation (GGA)-XC designed to improve the equilibrium properties of densely packed solids and their surfaces, making it a natural geometry-focused transfer test~\cite{perdew2008restoring}. 
Without retraining or fine-tuning, we replace only the PBE XC term in the Kohn--Sham potential with the PBEsol potential at each FNO iteration; PBEsol is also used for energy evaluation, while all other model inputs and inference settings remain unchanged. See SI.
The FNO based SCFs follow the resulting shift of the entire EOS: it predicts
$V_0=133.00$~\AA$^3$/cell and $B_0=45.15$~GPa, compared with
$132.20$~\AA$^3$/cell and $48.92$~GPa from self-consistent PBEsol. 
The corresponding errors of $0.61\%$ and $7.7\%$ are larger than for PBE but remain small despite the absence of any PBEsol data.

This functional transfer follows from the operator definition. 
The noninteracting output density is determined by the complete
local KS potential, not by the XC approximation used to construct it.
Potential-density pairs generated with different density-only XC
approximations are therefore samples of the same KS operator and can be mixed
in training without a functional label.
Similarly, at inference, the input KS potential to the model can be constructed using any density-dependent XC-potential approximation.
Fine-tuning would improve coverage, but the learning target need not be redefined.
Thus, even before learning the kinetic-energy output of the joint KS operator,
the density model can replace the full orbital-based periodic SCF loop involving 
accurate norm-conserving pseudopotentials.  
Density
and density-derived observables require no subsequent orbital calculation.
A conventional fixed-density solve is needed only when orbital-resolved spectra are needed, rather than at every $\mathbf{k}$ point of
every SCF iteration.

\subsection{Magnesium Dislocation}
\label{sec:results_Mg}

\begin{figure}[!htbp]
    \centering
    \includegraphics[width= \textwidth]{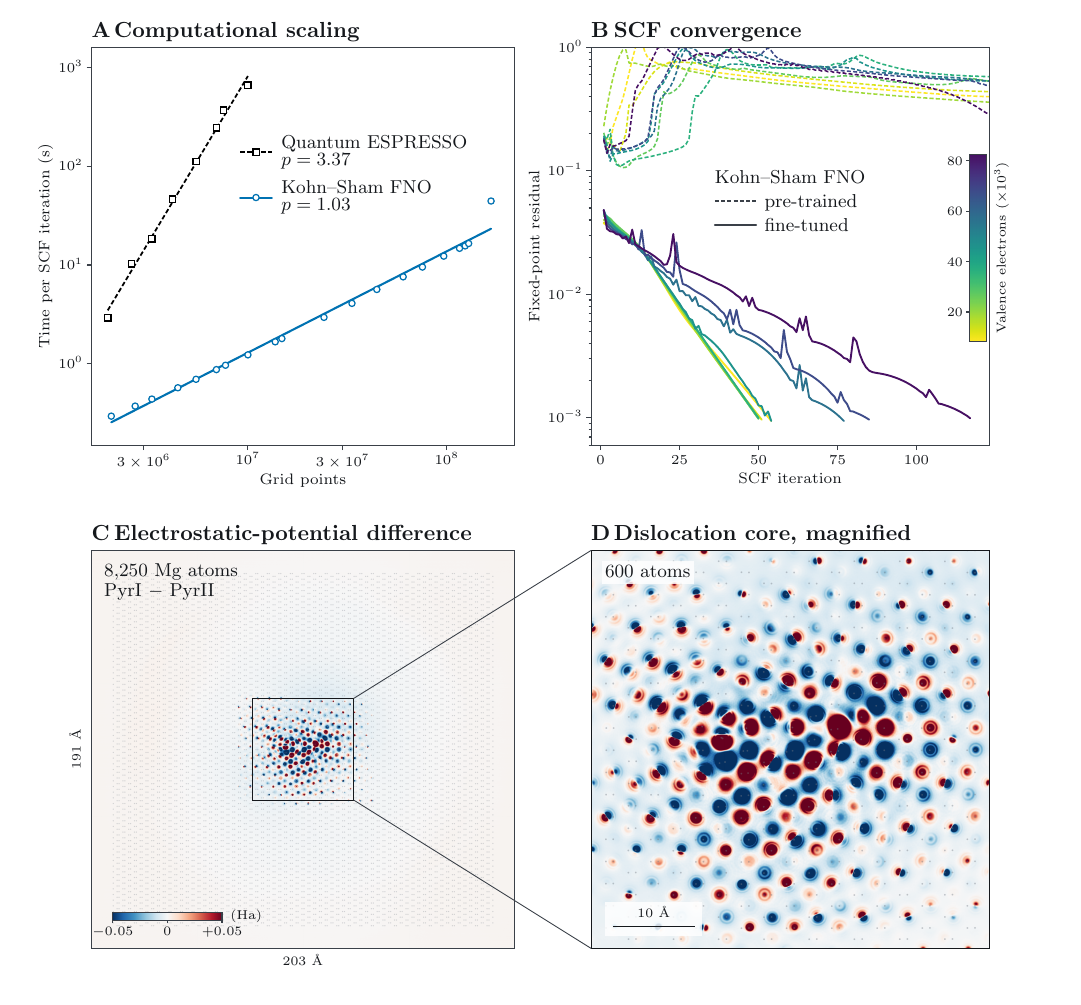}
    \caption{\textbf{Near linear-scaling density optimization for large metallic dislocation cells.}
    (A) Time per complete SCF iteration $t$ versus the number of real-space grid points $N_g$ on logarithmic axes.
    Symbols show Quantum ESPRESSO
(black squares) and Kohn--Sham FNO (blue circles) timings; lines are
empirical power-law fits $t\propto N_g^p$, where $p$ is the fitted scaling
exponent.
Quantum ESPRESSO calculations were run on 192-core CPU nodes, whereas the FNO used one
NVIDIA B300 GPU.
See SI.
(B) Fixed-point residuals for pyramidal-I and pyramidal-II dislocation
    cells using the base model (dashed), trained on the QM9 and MC3D training set in Fig.~\ref{fig:fig3_mols}A, and the model fine-tuned on Mg-defect data (solid; see SI); color denotes valence-electron count.
    (C) Difference in the electrostatic potential
    $v_{\mathrm{ext}}^{\mathrm{loc}}+v_{\mathrm H}[n]$ between the 8,250-atom
    pyramidal-I and pyramidal-II cells using Kohn--Sham FNO converged densities. 
    Grey dots denote the Mg atom locations from the pyramidal-I cell.
    The heat map shows the 99.5th percentile difference.
    (D) The dislocation core at full resolution. The region boxed in (C) contains 600 atoms and is magnified roughly 3.9x.
    Training, fine-tuning, and inference data generation, convergence, and timing calculations specific to these results: see SI.
    }
    \label{fig:fig5_Mg}
\end{figure}

Finally, we move from ideal periodic crystals to extended defects at realistic
length scales.
A dislocation breaks primitive translational symmetry and
produces a long-ranged elastic field, so first-principles calculations require
supercells containing thousands of atoms.
We consider $\langle c+a\rangle$ screw dislocations in Mg, the lightest structural metal.
Its limited ductility is linked to the relative stability and cross-slip of
competing pyramidal-I and pyramidal-II cores, which can be altered through
alloying~\cite{wu_mechanism_2016, wu_mechanistic_2018, das_intrinsic_2026}.
Accurately resolving their small energy difference therefore requires cell
sizes at which repeated orbital diagonalization becomes prohibitive.

This problem was used by the 2019 ACM Gordon Bell Prize finalist study of Das \textit{et al.} to demonstrate large-scale metallic DFT with finite elements (FE) based DFT-FE package
~\cite{das_fast_2019, motamarri2020dft}.
A full ground-state calculation for 6,164 Mg atoms
required 56 SCF iterations on 1,300 nodes of the Summit supercomputer
(7,800 NVIDIA V100 GPUs), whereas the largest system, containing 10,508
atoms and 105,080 electrons, used 3,800 nodes (22,800 GPUs) for one
SCF iteration.
To test the practical reach of the quasi-linear Kohn--Sham FNO,
we therefore attempt to converge the same class of metallic-defect SCFs on a single, modern high-memory GPU.

As a test, we began with the same pretrained Kohn--Sham FNO used for the molecular and
solid calculations above, without any defect-specific fine-tuning.
The
SCF trajectories immediately revealed that additional training was needed as 
the fixed-point residual rapidly increased for every tested cell and none of
the calculations converged [Fig.~\ref{fig:fig5_Mg}B].
This warning required no
reference density.
The predicted density was
inconsistent with the KS potential reconstructed from it, causing the next
model update to move farther from a fixed point.
Iterative prediction thus
provides a built-in reliability diagnostic that a direct ground-state prediction model would not. 
The residual is not a formal
error bound, and convergence alone does not certify agreement with PBE ground-state, but
SCF divergence immediately identifies a model that should not be trusted.

We therefore fine-tuned the same pretrained model on 1,203 Mg structures
containing 20-364 atoms. See SI.
The data combine
bulk and strained cells, surfaces, generalized stacking faults, and local
environments cut from pyramidal-I and pyramidal-II dislocation cores; the
architecture and inference procedure were unchanged.
The fine-tuned model
converges every tested dislocation cell to the same relative-$L^1$
fixed-point threshold of $10^{-3}$ used throughout this work
[Fig.~\ref{fig:fig5_Mg}B]. See SI.
The largest
calculation contains 8,250 Mg atoms and 82,500 valence electrons and runs on
one NVIDIA B300 GPU without reducing the real-space grid resolution. 
Their trajectories also contain occasional residual spikes, which we
attribute to numerical instabilities when applying the FNO on such large grids. 
Adaptive density mixing nevertheless returns each
trajectory to convergence. See SI.

Timing the complete SCF update confirms the expected near-linear scaling of the approach empirically [Fig.~\ref{fig:fig5_Mg}A]. 
Power-law fits against the number of
grid points give $p=1.03$ for the Kohn--Sham FNO, including the largest system,
compared with $p=3.37$ for Quantum ESPRESSO. 
The FNO result is consistent with
the expected $\mathcal O(N_g\log N_g)$ complexity of the complete FFT-based
SCF iteration. 
FNO timings used one B300 GPU, whereas Quantum ESPRESSO timings were linearly rescaled by core count to estimate execution on a 192-core AMD EPYC 9655 node (see SI); their absolute times are therefore not a
hardware-matched comparison. 
The upward deviation of the largest FNO point
likely reflects memory pressure near the capacity of the B300, which set the
maximum system size tested.

Because converging PBE reference densities for the full dislocation cells 
with Quantum ESPRESSO was computationally prohibitive, density errors were
evaluated on core-centered crops containing up to 528 Mg atoms, the largest
tractable size on nodes with 750~GB of memory. See SI.
Across the held-out
pyramidal-I and pyramidal-II cells, the mean FNO density error remains
$0.33$--$0.35\%$ with no systematic growth with size
[see SI]. 
At 8,250 atoms, where no reference is available,
the difference in the FNO-derived electrostatic potential (ESP)
$v_{\mathrm{ext}}^{\mathrm{loc}}(\mathbf{r})+v_{\mathrm H}[n](\mathbf{r})$ between the two core
structures remains localized around the dislocation
[Fig.~\ref{fig:fig5_Mg}C]. 
Although not an independent accuracy test, this
field shows that the large converged densities resolve distinct core
environments.

Figure~\ref{fig:fig5_Mg} tests only the density component of the joint KS
operator.
The converged densities can be passed to a single fixed-density orbital calculation, such as a DFT-FE calculation, or paired with a future forward kinetic-energy model to obtain total energies without repeating the full orbital SCF trajectory. 
Here, we establish stable self-consistency for an extended
metallic defect with more than $8\times10^4$ electrons on one GPU.

\section{Conclusion}
\label{sec:conclusion}

In this work, we identify the Kohn--Sham map as a complete orbital-free DFT
target and construct a domain-invariant, radially factorized Fourier neural operator
to learn its density component.
The model replaces one non-interacting orbital solution at an arbitrary SCF
iteration, rather than learning the inverse density-to-potential relation or
compressing the complete ground-state calculation into one prediction.
This factorization removes the repeated orbital diagonalization bottleneck from regular KS-DFT SCFs while retaining explicit Hartree and XC construction, density mixing, and self-consistent feedback.

Controlled comparisons using the same data and model backbone show that this
choice stabilizes density optimization relative to the inverse kinetic-
potential map and extrapolates more reliably than direct ground-state
prediction.
The same Kohn--Sham FNO drives molecular, semiconducting, and
metallic SCFs with transferable norm-conserving pseudopotentials as training reference.
It can learn
from truncated, unconverged reference trajectories, transfer from PBE to
PBEsol without retraining, and produce densities that recover KS-DFT quality electronic observables, spectra and equations of state.
After fine-tuning only on Mg cells
containing at most 364 atoms, the model converges dislocation-cell densities
containing up to 8,250 atoms and 82,500 valence electrons on one GPU, with an
empirical scaling of $\mathcal{O}(N^{1.03})$. 
The fixed-point residual additionally
provides a direct diagnostic when the learned operator is being applied
outside its reliable domain.

The present model learns only the density component of the joint Kohn--Sham
operator.
Orbital-resolved observables and total energies can already be obtained at KS-DFT accuracy with one fixed-density post-SCF calculation, but eliminating that
final orbital solve requires learning the corresponding forward kinetic-energy
(or finite-smearing free-energy) output.
Because both quantities are produced
by the same one-particle problem, we will pursue a variationally consistent framework for the complete map in future work. 
More broadly, these results show that machine learning can extend
electronic-structure calculations most effectively by replacing their
dominant repeated operation while preserving the physical iteration that
constructs and validates the ground state.

\section{Acknowledgements}
A. Anandkumar is supported in part by the Bren endowed chair, ONR (MURI grant N00014-18-12624), DARPA ExpMath HR0011, and by the AI2050 senior fellow program at Schmidt Sciences. S. Sharma was supported by the U.S. Department of Energy, Office of Science, Office of Basic Energy Sciences, Fuels from Sunlight Hub under Award Number DE-SC0021266. 
D. Khan acknowledges support from the Pritzker AI+Science fund and DARPA Biological Technologies HR0011. 
M. Hanisch is supported by the Kortschak Scholars Program.
E. Xie is supported through the Caltech Summer Undergraduate Research Fellowship.
D. Khan and M. Hanisch acknowledge discussions with Tommaso Chiarotti,
Valentin Duruisseaux, Chuwei Wang, Vignesh Bhethanabotla, Chenghan Li, and Garnet K. L. Chan. 
We acknowledge TACC, Schmidt Sciences and Caltech
HPC for computing resources.

\bibliographystyle{unsrtnat}
\bibliography{main}

\clearpage
\appendix

\end{document}